# Photochemical Upcycling of Ultrastrong Polyethylene Nanomembranes into Fibrous Carbon at Ambient Conditions

Yuexiang Sun[1,2], Xin Ma[1,2], Qiao Gu[3], and Ping Gao[2,3]*

**ABSTRACT**

The escalating global issue of plastic waste accumulation, specifically polyolefins, necessitates an urgent solution for upcycling these materials into beneficial compounds. Yet, achieving such upcycling without introducing carbon dioxide into the environment remains a formidable challenge. In this study, we demonstrate an eco-friendly approach for the photochemical conversion of ultrastrong, ultratransparent, and ultrathin polyethylene membrane into fibrous carbon nanomembrane at ambient conditions. The membrane was sputter-coated with platinum and cuprous oxide nanoparticles and exposed to simulated sunlight, resulting in a porous carbon membrane decorated with Pt nanoparticles. The new carbonized nanomembrane maintained the pristine membrane's morphology. The membrane exhibited high activity (2.11 mA/cm$^2$) for electrochemical ethanol oxidation with stability over 1000 cycles. This work holds significance for sustainable plastic waste management and the design of new polyolefin materials in a circular economy.

[1] Division of Emerging Interdisciplinary Areas, Interdisciplinary Program Office, The Hong Kong University of Science and Technology, Clear Water Bay, Kowloon, Hong Kong SAR 999077, China
[2] Thrust of Advanced Materials, The Hong Kong University of Science and Technology (Guangzhou), Nansha, Guangzhou 511400, China
[3] Department of Chemical and Biological Engineering, The Hong Kong University of Science and Technology, Clear Water Bay, Kowloon, Hong Kong SAR 999077, China
*Corresponding author. Email: kepgao@ust.hk

**INTRODUCTION**

Synthetic plastics have revolutionized our daily lives with their remarkable mechanical strength/weight ratios, exceptional chemical stability, and affordability[1]. However, these very qualities have also contributed to a global environmental crisis that affects ecosystems worldwide[2-4]. It is estimated that approximately 7 billion metric tons of plastic waste have been produced since the 1970s. Sadly, the majority of plastic waste has been either incinerated or sent to landfills, with less than 10% being recycled[5]. Incineration leads to significant carbon dioxide ($CO_2$) emissions, while landfilling results in a slow natural degradation process, leading to the persistent accumulation of plastic in the environment[6].

Plastic, fundamentally composed of carbon (C) and hydrogen (H), holds great potential as a valuable resource for the chemical industry. Consequently, the upcycling of waste plastic into useful materials, or plastic upcycling, emerges as an appealing solution to address the urgent problem of plastic waste[7-9]. This process aims to convert waste plastic into carbon-based materials and hydrogen, offering a promising avenue for sustainable utilization. By transforming plastic waste into valuable resources, we can mitigate the environmental impact and contribute to the transition towards a more circular economy.

A prospective resolution to this issue is the upcycling of plastics through carbonization, a process that has garnered substantial interest of late for its ability to convert waste into valuable carbon-based materials[10,11]. This process involves the decomposition of plastics or polymers at elevated temperatures or pressures in an oxygen-free environment, culminating in the formation of carbon-rich materials. The products of carbonization, compared to those of other recycling processes, are highly controllable and economically beneficial, with reduced $CO_2$ emissions[12]. The plastics or polymer can be carbonized to porous carbon[13-15], carbon fiber[16-18], carbon nanotube[19-22], etc. These materials have a wide range of applications, including energy storage[23], catalysis[24], and environmental remediation[25]. However, the existing methods developed so far for plastic upcycling have been energy-intensive, requiring high temperatures or high-pressure conditions. This challenge arises from the inherent chemical stability of synthetic polymers, which makes them difficult to transform easily. Therefore, there is an urgent need for an innovative, cost-effective, and energy-efficient approach to address the upcycling of plastics.

In this study, we report a novel approach for the low energy upcycling of polyolefin materials at ambient conditions. Specifically, we have discovered that the porous and ultrastrong ultra-high molecular weight polyethylene (UHMWPE) nanomembrane developed in our laboratory[26] can be converted into fibrous carbon nanomembranes under ambient conditions by a facile photocatalytic process. To achieve this, we simply coated the nanomembrane with platinum (Pt) and copper (I) oxide ($Cu_2O$) nanoparticles and exposed the membranes to simulated sunlight. Remarkably, after only 10 minutes of irradiation under 5 times the intensity of sunlight, the high-strength membrane was converted into nanofibrous carbon with supported Pt nanoparticles. The carbonized Pt-PE sample was then tested for fuel electrochemical ethanol oxidation (EOR) and was observed to exhibit high activity and stability. This pioneering discovery offers significant potential for the fabrication of new sustainable polymeric materials that can contribute to a circular economy.

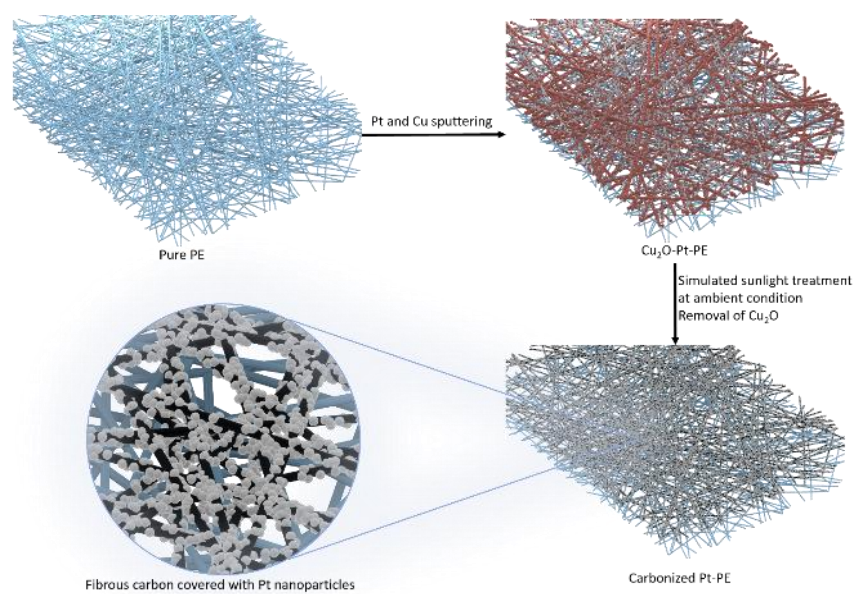

**Figure 1.** The preparation process of the PE membrane decorated with Pt and $Cu_2O$ nanoparticles and the carbonization of PE into fibrous carbon.

**RESULTS**

A schematic diagram illustrating of the two-step approach for the transformation of UHMWPE nanomembrane into Pt nanoparticle/fibrous carbon nanocomposite membrane is depicted in Figure 1. The pristine UHMWPE nanomembrane was first coated with a monolayer of $Cu_2O$ and Pt nanoparticles using magnetron sputter coating and Cu oxidation, forming a $Cu_2O$-Pt-PE composite. This is then followed by irradiating the membrane in the presence of water using a simulated sunlight lamp source at ambient condition for 10 minutes (see Methods for the detailed preparation process). After the washing away of the Cu species, we obtained a carbonized Pt-PE membrane. Figure 2a presents an optical image of the synthesized $Cu_2O$-Pt-PE nanomembrane before carbonization. The sample is partially transparent because of the highly transparent and porous UHMWPE nanofilm (~200 nm in thickness) substrate[26] and the ultrathin deposition of the Pt and Cu (~10 nm in thickness). Figure 2b showed the sample after carbonization with the intact membrane structure. The atomic force microscopic (AFM) images in Figure 2c-2e present the structure evolution from pristine PE, $Cu_2O$-Pt-PE and eventually to carbonized Pt-PE. The pristine PE (Figure 2c) is a fibrous nanomembrane with fiber diameters of the order 20 nm[26]. The $Cu_2O$-Pt-PE (Figure 2d) presents a nanoparticle coated PE fibrous structure, with Pt and $Cu_2O$ nanoparticles uniformly distributed along the PE nanofibers. The X-ray diffraction (XRD) pattern of pure PE, Pt-PE, and $Cu_2O$-Pt-PE are displayed in Figure S1. The peaks at 2θ of 21.6° and 24.0° can be ascribed to the (110) and (200) facets of the orthorhombic UHMWPE membrane, respectively. The two relatively weak and broad peaks at 2θ of 39.8° and 36.4° can be assigned to Pt (111) (JCPDS Card No. 04-0802) and $Cu_2O$ (111) (JCPDS Card No. 99-0041). It is interesting to note that the carbonized Pt-PE (Figure 2e) maintained the interpenetrating fibrous nanostructure analogous to that of the $Cu_2O$-Pt-PE before carbonization. Figure 2f showed the scanning electron microscope image of the carbonized Pt-PE as well as the energy dispersive spectrometer (EDS) mapping result of the Pt element. The Pt particles are uniformly distributed on the fiber.

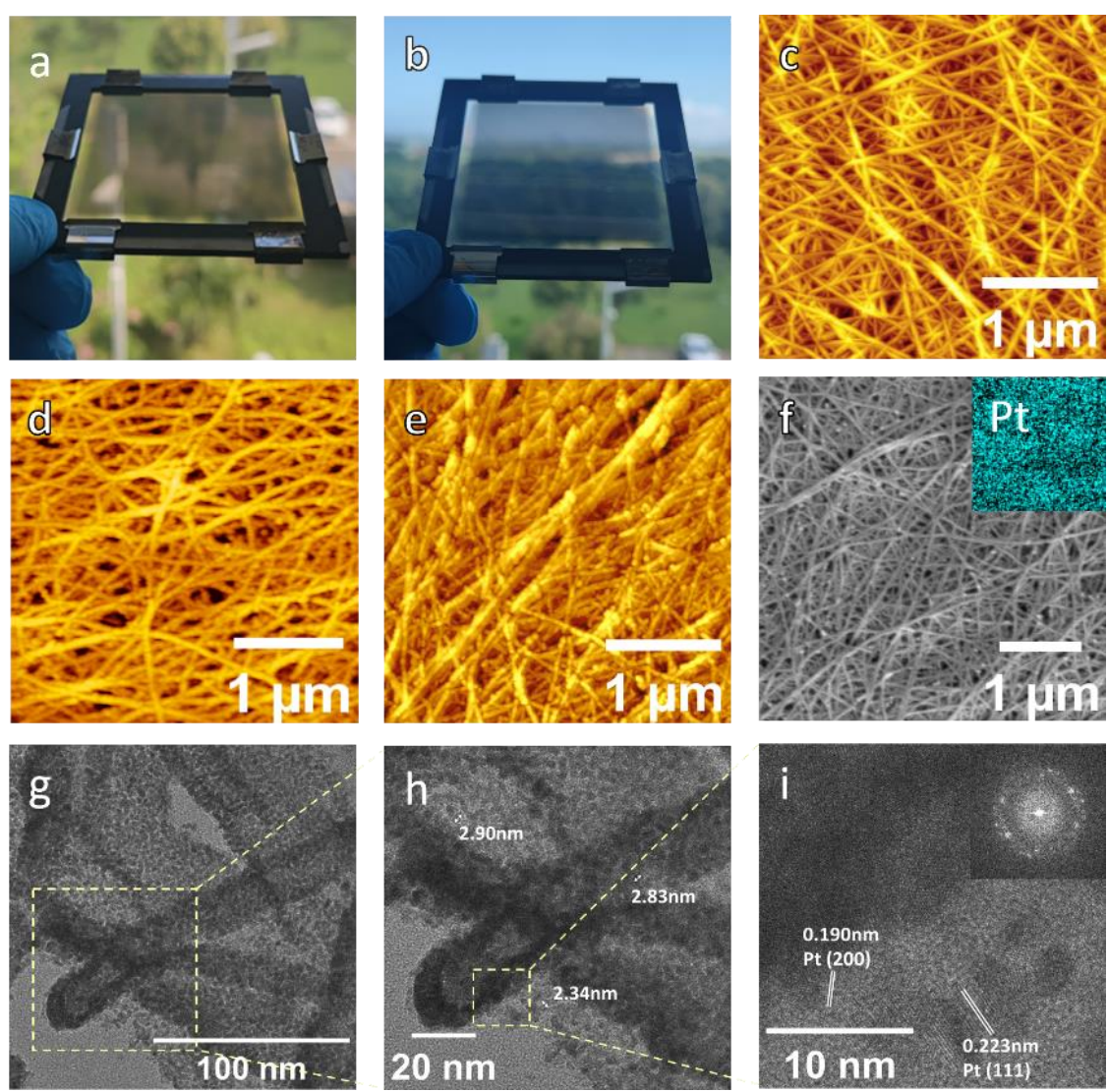

**Figure 2.** The optical image of **a)** $Cu_2O$-Pt-PE and **b)** carbonized Pt-PE. AFM image of **c)** Pure UHMWPE membrane, **d)** $Cu_2O$-Pt-PE **e)** carbonized Pt-PE. **f)** SEM image of carbonized Pt-PE and the EDS result of the Pt. **g)** and **h)** and **i)** HRTEM image and its FFT image of carbonized PE.

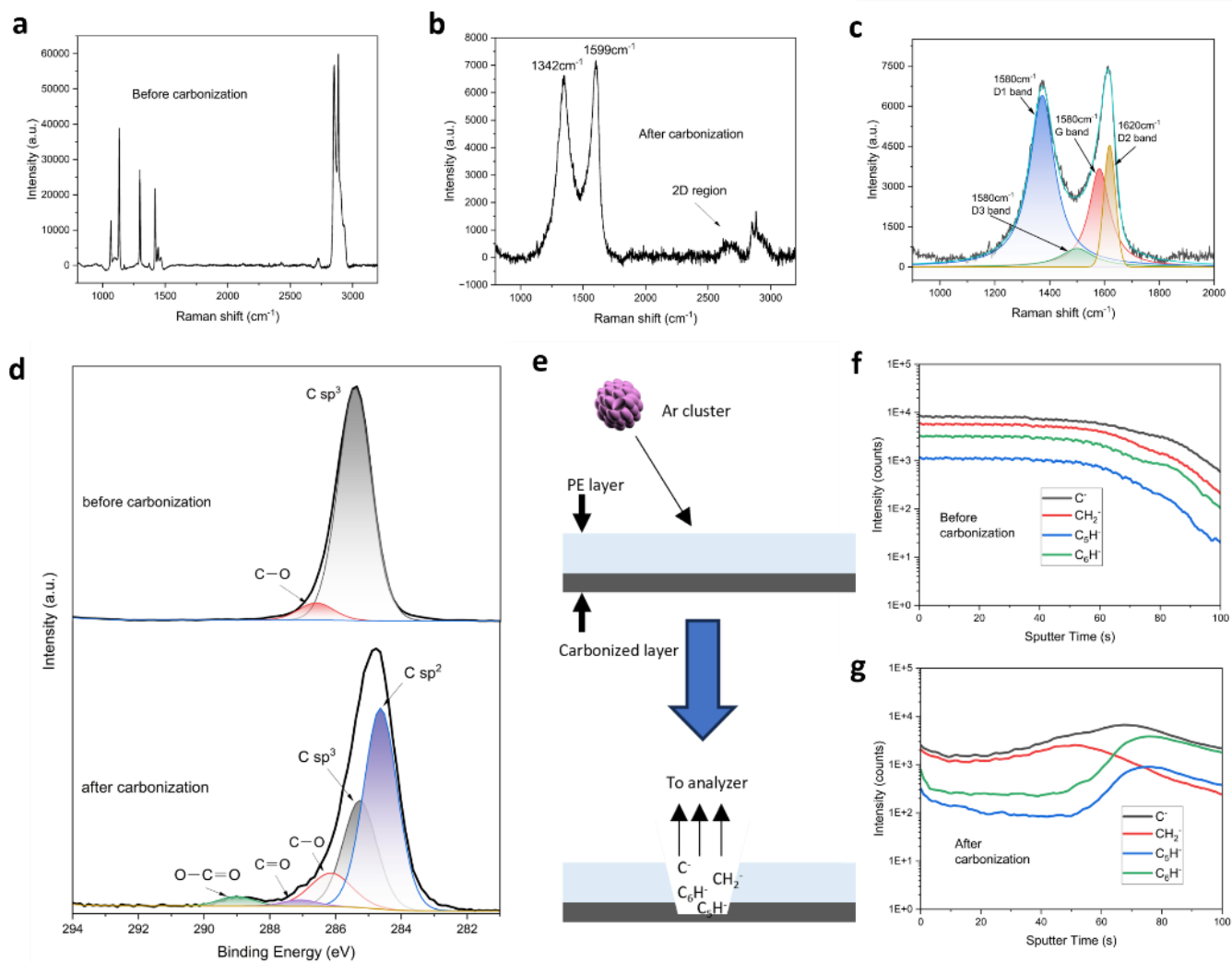

**Figure 3.** Raman scattering spectra of **a)** Pt-PE before carbonization and **b)** Pt-PE after carbonization. **c)** The Raman peak deconvolution results with Five-peak model. **d)** The C1s XPS results of PE before and after carbonization. **e)** The schematic illustration of the Ar cluster etching during the ToF-SIMS test. ToF-SIMS depth profile of **f)** Pt-PE before carbonization and **g)** Pt-PE after carbonization.

To elucidate the carbonization of the PE nanomembrane, we performed transmission electron microscopy (TEM) analysis on the sample exfoliated from the carbonized Pt-PE membrane by ultrasonication. Interestingly, the carbonized particles also retained the fibrous structures of the pristine PE membrane as shown in Figures 2g and 2h. Using fast Fourier transform (FFT) and d-spacing analysis of the fringes (Figure 2i), we observed the exposed surfaces of the Pt particles were predominantly (111) and (200) facets. Besides, these TEM images show that Pt nanoparticles are quite uniform in size (~3 nm) and distributed uniformly along fiber axial direction. It is remarkable that these Pt nanoparticles did not coagulate during the intensive exfoliation process of 24 hours sonication. This implicates there exists strong attractive force at the interface between the Pt nanoparticles and the newly carbonized carbon fibers.

Raman scattering spectroscopy is the most powerful tool for carbon material characterization. The Pt-PE sample before and after carbonization was tested using a 532 nm laser, and results are shown in Figure 3a and 3b. The pure PE sample exhibited a typical UHMWPE pattern and no sign of any carbonaceous materials with $sp^2$ hybridizations. In contrast, for the carbonized sample, the Raman spectra exhibited two board bands at Raman shift of around 1342 cm$^{-1}$ and 1599 cm$^{-1}$. The Raman peaks were deconvoluted according to Sadezky et al.'s five-peak model[27] as seen in Figure 3c. The spectra deconvoluted using Lorentzian peaks near Raman shift of 1580, 1350, 1620, and 1200 cm$^{-1}$ and one Gaussian peak at 1500 cm$^{-1}$, representing the G, D1, D2, D4 and D3 band, respectively. The G band originates from the ideal graphitic lattice, while the D1 and D2 bands represent the disordered graphitic lattice. Clearly, the carbonized PE primarily consists of the graphitic phase. The presence of weak D3 band at 1500 cm$^{-1}$ suggests it also contained a small amount of amorphous carbon. The missing D4 band at 1200 cm$^{-1}$ indicates that the residue polyenes impurities were not detected[27]. Interestingly, the Raman spectra also revealed a weak 2D band at approximately 2700 cm$^{-1}$. Overall, the carbonized PE exhibited high graphitic crystallinity.

The transformation of polyethylene to carbon material is a process in which $sp^3$ carbon atoms are converted into $sp^2$ carbon atoms. In order to investigate the chemical status of this transformation, we performed X-ray photoelectron spectroscopy (XPS) analysis, as depicted in Figure 3d. The XPS spectra of C1s provided solid evidence for the carbonization process, as the carbon peak shifted to a lower binding energy area, from 285.2 eV to 284.5 eV. In addition to the changes in carbon chemical status, the peak deconvolution revealed the emergence of $sp^2$ carbon, which is a key indicator of the carbonization process. The appearance of C-O, C=O and O-C=O peaks suggested the occurrence of partial oxidation at the edge or surface of the carbon species We also examined the chemical status of the Cu and Pt species before and after the carbonization process as shown in Figure S2. The Cu presented some 2+ chemical status after the process which can be attributed to the following chemical reaction:

$$Cu_2O+H_2O+O_2 \rightarrow Cu(OH)_2$$

The Pt species demonstrated a very stable chemical status before and after carbonization.

To further corroborate the carbonization process, we also conducted depth profiling analysis under Time-of-Flight Secondary Ion Mass Spectrometry (ToF-SIMS). The carbonized side of the sample was affixed to the Si wafer. The sample was etched by Ar clusters from the uncarbonized side as illustrated in Figure 3e. For the uncarbonized PE sample (Figure 3f), the unsaturated carbon groups ($C_5H^-$ and $C_6H^-$) did not have any obvious change when the etching depth reached the Si and sample interface. But for the carbonized sample, these groups intensity increased significantly as seen in Figure 3g. The 3D ToF-SIMS distribution also showed the concentration profile of carbonized PE (Figure S3). As the etching process neared completion, the $C^-$ (represented by red particles) dominated in comparison to the $CH_2^-$ ions (represented by blue particles), which is indicative of dehydrogenation. Moreover, the $C_5H^-$ distribution was found to be consistent with the previously obtained results. Depth profiling analysis was also conducted under XPS to investigate the changes in chemical status of the carbon species in the carbonized Pt-PE sample. The sample was tested from the PE side, like the ToF-SIMS experiment. The corresponding XPS spectra with different etching depth before and after carbonization are presented in Figure S4. For the PE sample prior to carbonization, no significant changes in the chemical status of the C1s core level were observed. However, after carbonization, the PE sample exhibited an increase in sp2 carbon species as the etching depth increased. This observation suggests that the PE fiber coated with Pt underwent carbonization. The above depth profile measurements by ToF-These results show that the carbonization occurred at the photocatalyst/PE interface without causing significant temperature to increase in the substrate for had the temperature been above that of the melting temperature of the PE substrate, we would have observed fiber morphology changes.

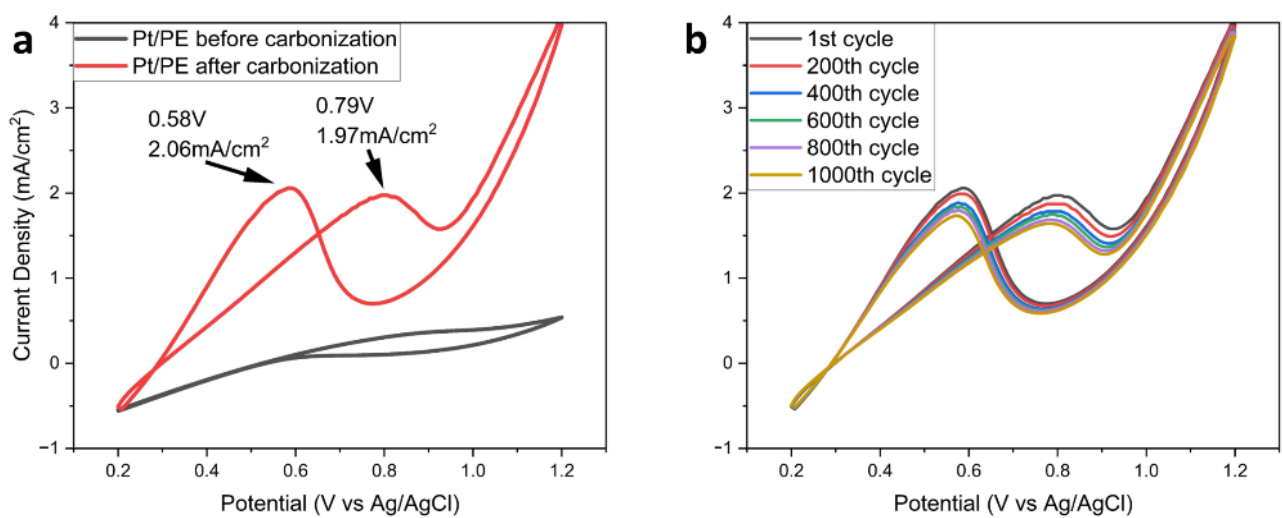

**Figure 4. a)** EOR curves of the carbonized Pt/PE and before carbonized Pt/PE. **b)** The stability test of EOR for 1000 cycles.

To demonstrate the functionality of the carbonized Pt/carbon nanomembrane, we characterized its catalytic performance for use as an electrocatalyst for ethanol oxidation reaction (EOR). The pristine Pt-PE sample before carbonization was also tested for comparison. The tests were conducted by a three-electrode system shown in Figure S5. Typical EOR curves of both samples are depicted in Figure 4a. The specific activity of the carbonized sample was 2.11 mA/cm$^2$, which is approximately 10 times higher compared with the Pt-PE before carbonization. The forward and backward peak position of carbonized Pt-PE sample are 0.84 V and 0.58 V (vs Ag/AgCl) while the Pt-PE before carbonization are 0.95 V and 0.67 V (vs Ag/AgCl), which suggested the catalytic activity enhancement after carbonization. The carbonization process not only improved the conductivity of the sample but also provided a Pt-carbon structure, which has been demonstrated to effectively enhance the catalytic performance by stabilizing the Pt nanoparticles for alcohol oxidation reactions[28]. The stability of the carbonized PE sample was evaluated by performing 1000 continuous cycles of EOR. The specific activity of the sample was maintained at 81% of its initial value (Figure 4b), indicating the high durability and long-term stability of the carbonized sample. The TEM image after the test is shown in Figure S6.

The proposed carbonization process mechanism is depicted in Figure 5. The Pt nanoparticles are believed to weaken the C-H bond[29]. Concurrently, the Cu$_2$O semiconductor generates the electron-hole pairs with the illumination of the light. The plasmonic effect of the interconnected Pt nanoparticles enhances the local light intensity, acting as an electron donor[30] and augmenting the oxidative properties of the system. Consequently, the hole from the Cu$_2$O valence band (VB) transfers to Pt, initiating the extraction of hydrogen (H) from the system. Moreover, $H_2O$ molecules draw out an additional H from the PE substrate, resulting in the formation of carbon double bonds. The electrons present in the conduction band (CB) then reduce dissolved $O_2$ in water, producing hydroxide ions ($OH^-$). Subsequent cycloaddition and dehydrogenation reactions further transform the intermediates, ultimately leading to the formation of distinct carbon species. To assess the surface plasmon resonance effect of Pt nanoparticles, we employed wave optics analysis using COMSOL Multiphysics modeling, with the results presented in Figure S7. The magnitude of electric filed intensity ($|E/E_0|$) mapping was simulated to represent the light intensity for a bunch of 30 nm Pt nanoparticles with different gaps (10nm, 5nm and 0) under an incident light from top side. The nanoparticles string with 0 gap was used to represent our Pt nanoparticles located on PE fibers. The electric field intensity of the contiguous Pt nanoparticles string was found to be significantly higher at the interface between the nanoparticles (~88 times) than that of the individual, isolated nanoparticles with gaps of 10 nm, and around 38 times higher than the nanoparticles with gaps of 5 nm. This indicates that the plasmonic effect of the Pt nanoparticles on PE fiber is able to enhance the light intensity significantly and finally initiate the carbonization process.

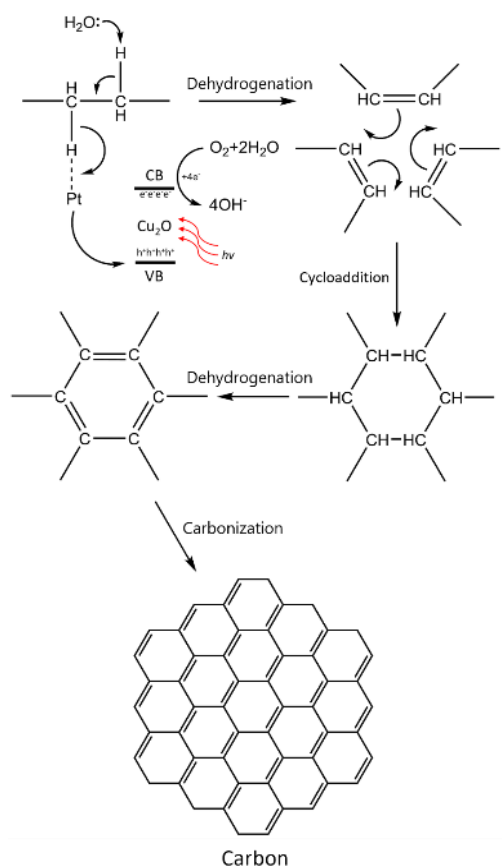

**Figure 5.** The proposed mechanism of PE carbonization process.

## CONCLUSION

Conventional plastic carbonization upcycling methods often necessitate high temperatures and/or pressures, resulting in substantial energy consumption. Our research proposes a low-energy, ambient-condition carbonization process for upcycling UHMWPE membranes into fibrous carbon, offering a promising and sustainable solution to the pervasive issue of plastic pollution. Our method developed here, while particularly suitable for nanofibrous PE membranes, can also be extended into other polymer films. This innovative method employs a straightforward photocatalytic process to transform waste plastic into valuable carbon-based materials. Additionally, the carbonized Pt-PE samples exhibit significant potential for electrochemical ethanol oxidation, suggesting their potential applicability in fuel cell technology. This novel upcycling approach not only addresses the environmental challenges associated with plastic pollution but also holds the potential to inspire the development of new upcycling strategies in the future, leading to the development of more environmentally sustainable polymeric materials.

## METHODS

### Materials

The ultrahigh molecular weight polyethylene (UHMWPE) resin (GUR 4022) with weight average molecular weight of 5 million Dalton was purchased from Celanese China Ltd. All other chemicals were purchased from Merklin without any further purification before all experiments. The supply gas used in this work was 99.999% ultra-high purity grade Argon (Ar). The deionized (DI) water was obtained using a Millipore Milli-Q system (18.2 MΩ·cm).

### Sample preparation

The free-standing porous UHMWPE ultrathin membrane was prepared using a method invented in our laboratory[26].

The Pt-PE membrane was prepared by depositing 10 nm Pt with the thin film sputtering system (Explorer, Denton vacuum, LLC). The sputtering rate was controlled at 20 nm/min. Cu-Pt-PE membrane was prepared by depositing 10 nm Cu on the previous Pt-PE sample. The sputtering rate was controlled at 30 nm/min. The oxidation of the Cu-Pt-PE sample was carried out at 70°C for 48 hours with the presence of $O_2$ to form $Cu_2O$-Pt-PE.

### Photocatalytic conversion of PE to carbon and preparation of carbonized Pt-PE

The prepared $Cu_2O$-Pt-PE sample placed in a quartz petri dish and soaked in water and illuminated with a simulated sunlight system (PLS-SXE300D, Beijing Perfectlight Technology Co.,Ltd) equipped with a light homogenizer. The distance between the sample and light source was 15 cm. The light intensity illuminated on the sample was $500\pm20$ mW/cm$^2$, which was tested by a light meter. The total reaction time was 10 minutes. Then the sample was firstly washed with 1M $FeCl_3$ solution to wash away the residue Cu species and then washed with both DI water and ethanol for three times for further characterization and electrochemical test.

### Material characterization

The X-ray diffraction (XRD) test was conducted by an X'pert Pro (PANalytical) diffractometer with Cu-Kα radiation (λ = 1.54178 Å) at a scan rate of 3°/min. The atomic force microscope (AFM) image was conducted by the Scanning Probe Microscope (Dimension ICON, Bruker) with contact mode. The scanning electron microscopy and the energy dispersive spectrometer test was conducted by the HITACHI 8320. The Raman scattering spectra was conducted by a confocal Raman microscope (inVia-Qontor, Renishaw). The Time-of-Flight Secondary Ion Mass Spectrometry (ToF-SIMS) test was performed with IonTof M6 (IonTof). The X-ray photoelectron spectroscopy test was conducted by PHI VersaProbe4 (ULVAC-PHI). The transmission electron microscopy (TEM) images were captured by the Talos F200C (Thermo Fisher Scientific), with an acceleration voltage of 200 kV. The TEM sample was prepared by sonicate the carbonized Pt-PE sample for 24 hours in ethanol.

**Electrochemical ethanol oxidation test**

The fuels electrochemical ethanol oxidation test was conducted by CHI760 electrochemical workstation. A three-electrode system was used for the whole test. A Pt clamping electrode was used as working electrode to hold the carbonized Pt-PE membrane electrode. The Ag/AgCl electrode was used as the reference electrode. A graphite electrode was used as the counter electrode. The ethanol oxidation reaction was conducted in a 0.5M $H_2SO_4$ + 1.0M ethanol solution electrolyte. Before the test, the cell was supplied with Ar gas with the rate of 10 sccm for 30 min to evacuate the $O_2$ within the system. The scan rate was 50 mV/s. The stability test was conducted with the same method for 1000 cycles continuously.

**COMSOL Multiphysics modeling**

The modeling of the surface plasmonic effect of the photocatalyst was performed by COMSOL Multiphysics (version 5.6). The electromagnetic field was simulated by the optical module. The model was surrounded by the perfectly matched layer in three directions. A normal incident light with a wavelength of 500nm radiated from the top side. We used 30-nm Pt nanoparticles string to simulate the Pt-PE sample. The gaps between the nanoparticles were 10 nm, 5 nm and 0 nm. The refractive index of Pt was calculated from the Drude model:

$$n_{Pt} = \sqrt{\varepsilon_m \mu_m}$$

$$\varepsilon_m = 1 - \left[\frac{\omega_p^2}{\omega(\omega + i\omega_c)}\right]$$

where $n_{Pt}$ is the refractive index of Pt, $\varepsilon_m$ is the relative permittivity of metal, $\mu_m$ is the relative permeability of metal, $\omega$ is the frequency of the incident light, $\omega_p$ is the plasma frequency and $\omega_c$ is the collision frequency. For Pt, $\mu_m = 1.000265$, $\omega_p = 7.8 \times 10^{15}\ rad/s$ and $\omega_c = 1.05 \times 10^{14}\ rad/s$ [31].

**Acknowledgements**

The authors gratefully acknowledge the financial supports from Guangzhou Municipal Government and the Research Institute of Tsinghua, Pearl River Delta (grant number RITPRD23EG01). The authors also wish to acknowledge the technical assistances from the technical staff of Materials Characterization and Preparation Facility in both Clear Water Bay and Guangzhou Campus of HKUST.

# Photochemical Upcycling of Ultrastrong Polyethylene Nanomembranes into Fibrous Carbon at Ambient Conditions

Yuexiang Sun[1,2], Xin Ma[1,2], Qiao Gu[3], and Ping Gao[2,3]*

**Supporting Information**

[1] Division of Emerging Interdisciplinary Areas, Interdisciplinary Program Office, The Hong Kong University of Science and Technology, Clear Water Bay, Kowloon, Hong Kong SAR 999077, China
[2] Thrust of Advanced Materials, The Hong Kong University of Science and Technology (Guangzhou), Nansha, Guangzhou 511400, China
[3] Department of Chemical and Biological Engineering, The Hong Kong University of Science and Technology, Clear Water Bay, Kowloon, Hong Kong SAR 999077, China
*Corresponding author. Email: kepgao@ust.hk

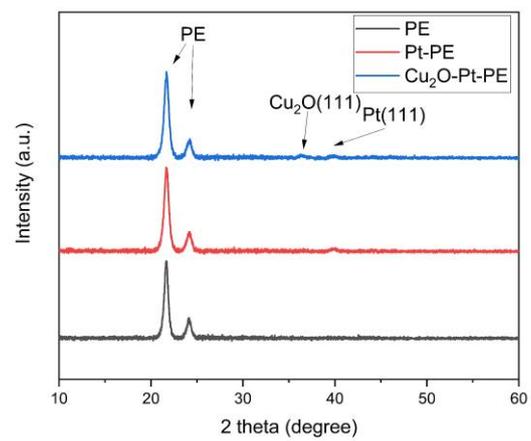

**Figure S 1.** The XRD pattern of pure PE, Pt-PE, Cu$_2$O-Pt-PE.

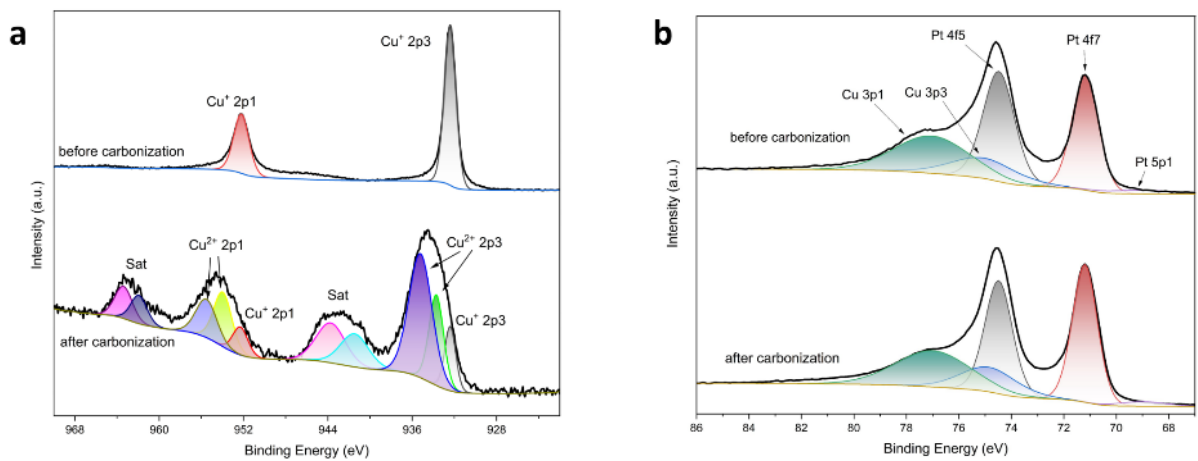

**Figure S 2.** XPS results of Cu and Pt species before and after carbonization.

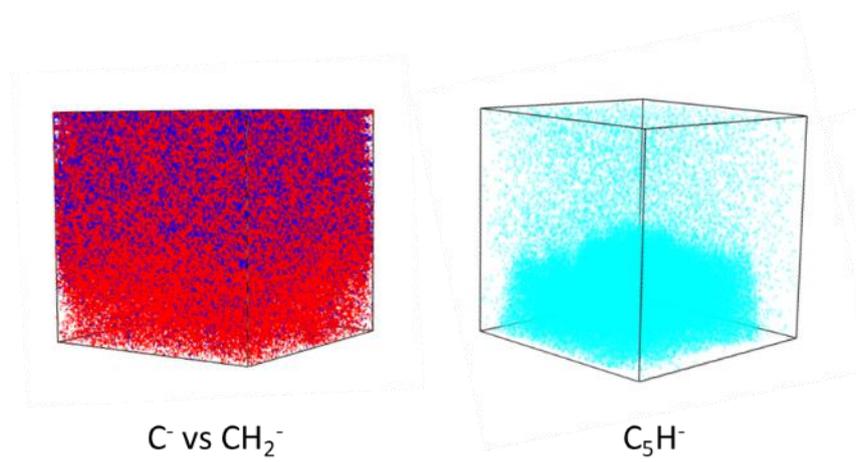

**Figure S 3.** The 3D ToF-SIMS distribution of the C⁻ (red) vs CH$_2^-$ (blue), C$_5$H⁻ (celeste) of the carbonized PE sample.

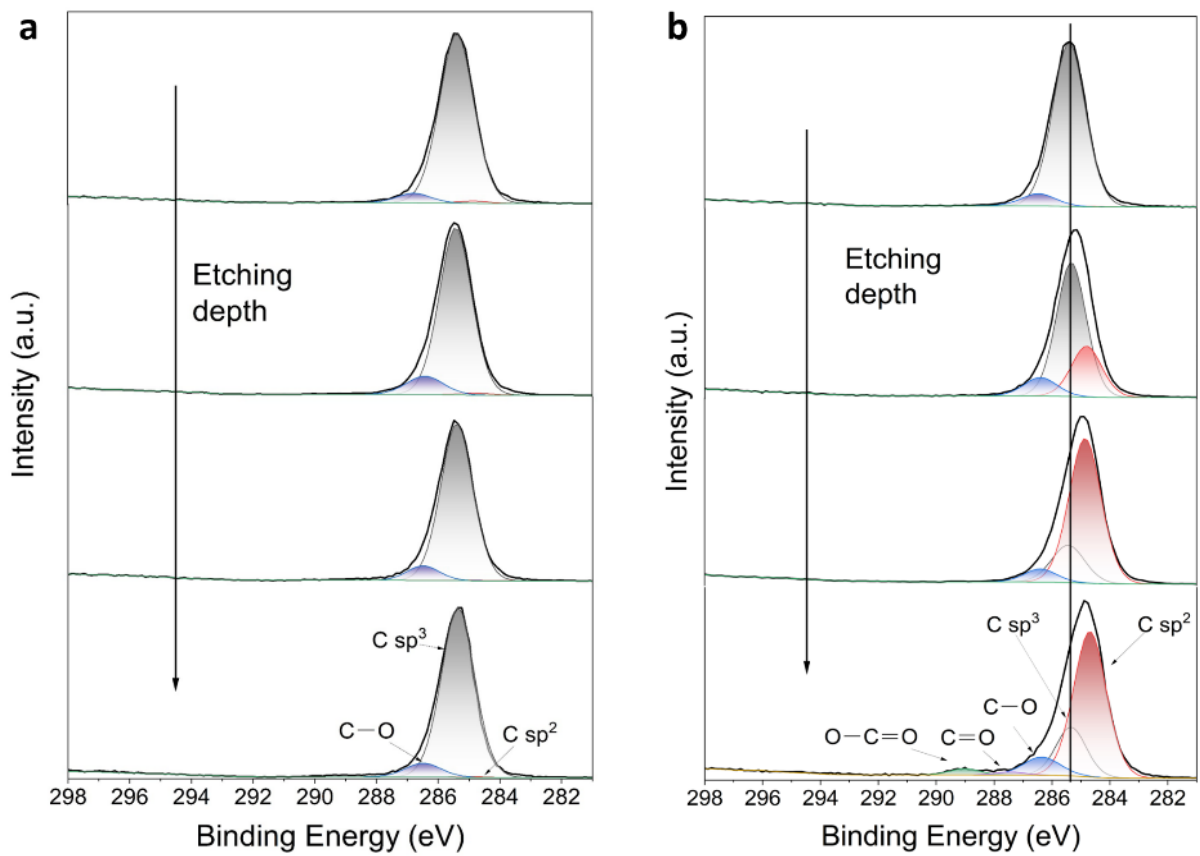

**Figure S 4.** The XPS results of C1s depth profile of PE a) before and b) after carbonization.

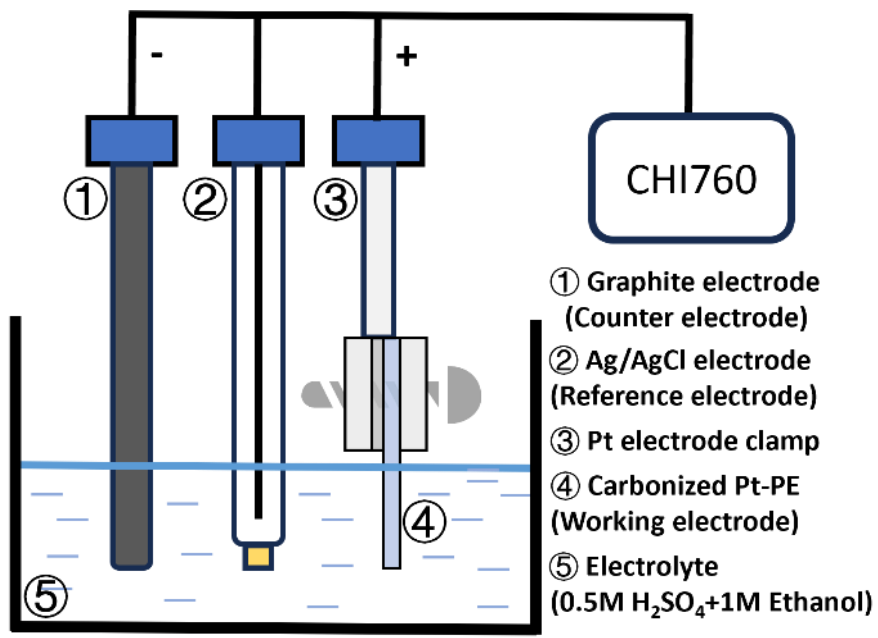

**Figure S 5.** The schematic illustration of electrochemical ethanol oxidation setup.

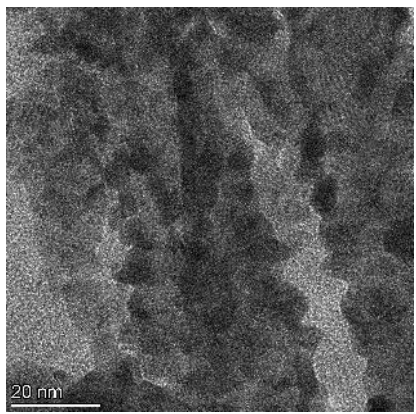

**Figure S7.** TEM image after the electrochemical test.

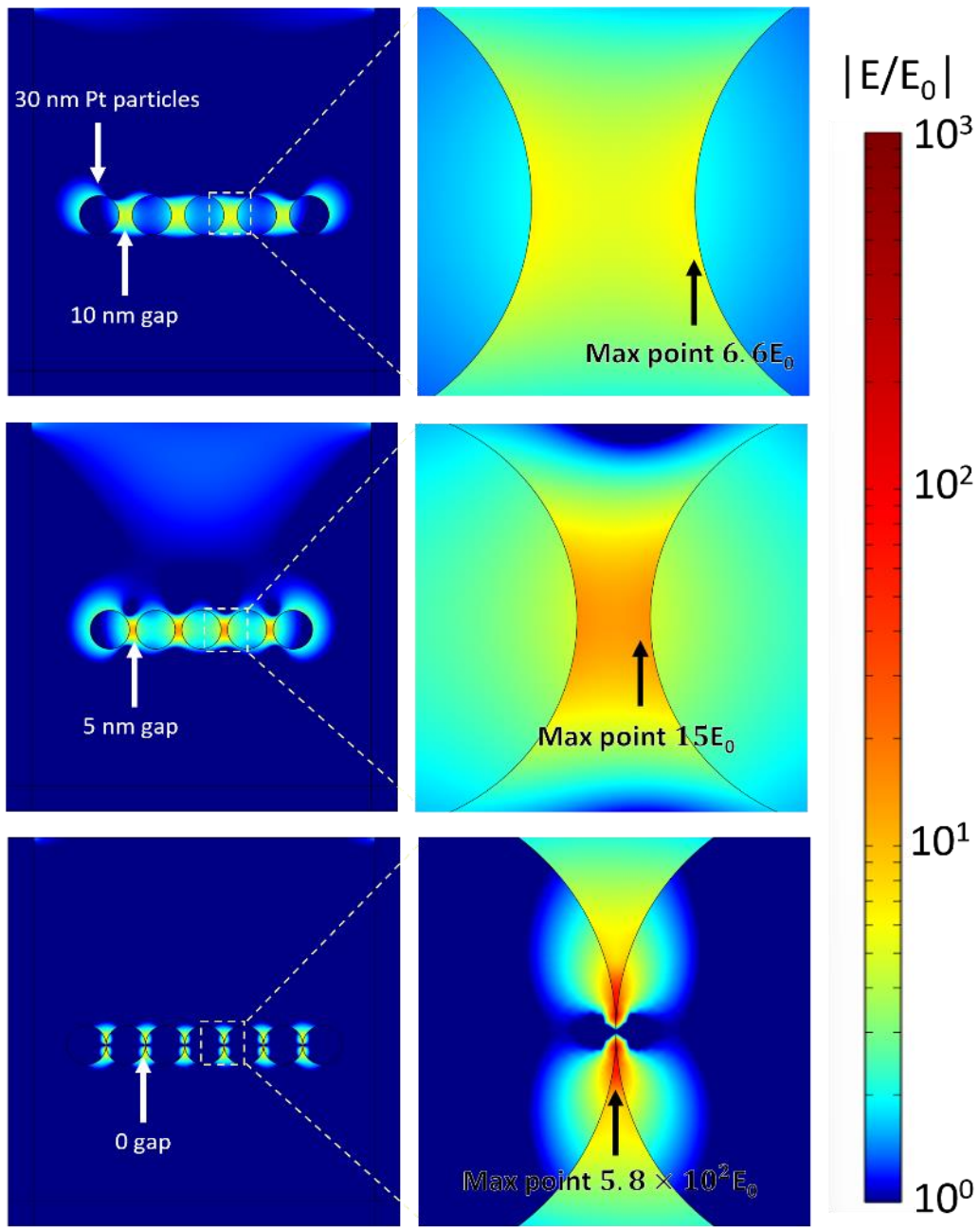

**Figure S7.** The electronic filed simulation of plasmonic effect of separated and conjunct Pt nanoparticles.